\documentclass{article}
\usepackage{ijcai23}

\usepackage{times}
\usepackage{soul}
\usepackage{url}
\usepackage[hidelinks]{hyperref}
\usepackage[utf8]{inputenc}
\usepackage[small]{caption}
\usepackage{graphicx}
\usepackage{amsmath}
\usepackage{amsthm}
\usepackage{booktabs}
\usepackage{algorithm}
\usepackage{algorithmic}
\usepackage[switch]{lineno}

\usepackage{color}
\usepackage{amssymb}
\usepackage{multicol}
\usepackage{multirow}
\usepackage{epsfig}
\usepackage{subfig}
\usepackage{xfrac}
\usepackage{bbding}
\usepackage{fontawesome5}

\title{Spatially Covariant Lesion Segmentation}

\author{
Hang Zhang$^1$
\and
Rongguang Wang$^2$ \and
Jinwei Zhang$^1$ \and
Dongdong Liu$^3$ \and
Chao Li$^1$ \And
Jiahao Li$^1$ 
\affiliations
$^1$Cornell University\\
$^2$University of Pennsylvania\\
$^3$New York University
\emails
\{hz459\}@cornell.edu,
rgw@seas.upenn.edu,
\{jz853\}@cornell.edu,
ddliu@nyu.edu
\{cl2527,jl3838\}@cornell.edu,
}

\begin{document}

\maketitle

\begin{abstract}

Compared to natural images, medical images usually show stronger visual patterns and therefore this adds flexibility and elasticity to resource-limited clinical applications by injecting proper priors into neural networks.
In this paper, we propose spatially covariant pixel-aligned classifier (SCP) to improve the computational efficiency and meantime maintain or increase accuracy for lesion segmentation.
SCP relaxes the spatial invariance constraint imposed by convolutional operations and optimizes an underlying implicit function that maps image coordinates to network weights, the parameters of which are obtained along with the backbone network training and later used for generating network weights to capture spatially covariant contextual information. 
We demonstrate the effectiveness and efficiency of the proposed SCP using two lesion segmentation tasks from different imaging modalities: white matter hyperintensity segmentation in magnetic resonance imaging and liver tumor segmentation in contrast-enhanced abdominal computerized tomography.
The network using SCP has achieved 23.8\%, 64.9\% and 74.7\% reduction in GPU memory usage, FLOPs, and network size with similar or better accuracy for lesion segmentation. 

\end{abstract}

\section{Introduction}

Convolutional neural networks (CNNs) have been the dominant approach across a variety of medical image computing applications such as magnetic resonance imaging (MRI) reconstruction~\cite{zhang2020fidelity,zhang2021efficient}, brain lesion segmentation ~\cite{ZHANG2021102854,zhang2019rsanet}, disease identification~\cite{KHOSLA2019651,zhang2022qsmrim}, and low-dose CT denoising~\cite{wang2022ctformer,fan2019quadratic}.
Two main properties \cite{elsayed2020revisiting} of CNNs are believed to be the key to their success: 1) The local receptive field increases as the network goes deeper \cite{huang2017densely,he2016deep}, providing richer contextual information for down-stream tasks; 2) Spatially invariant convolution filters regularize the network training, serving as a good inductive bias for grid-based image data \cite{simoncelli2001natural}.
However, medical images scanned from human body exhibit stronger patterns than natural images and therefore the question of whether the spatial invariance is overly restrictive in certain clinical applications remains to be answered.

\begin{figure}[!t]
	\centering
        \includegraphics[width=\columnwidth,height=0.345\columnwidth]{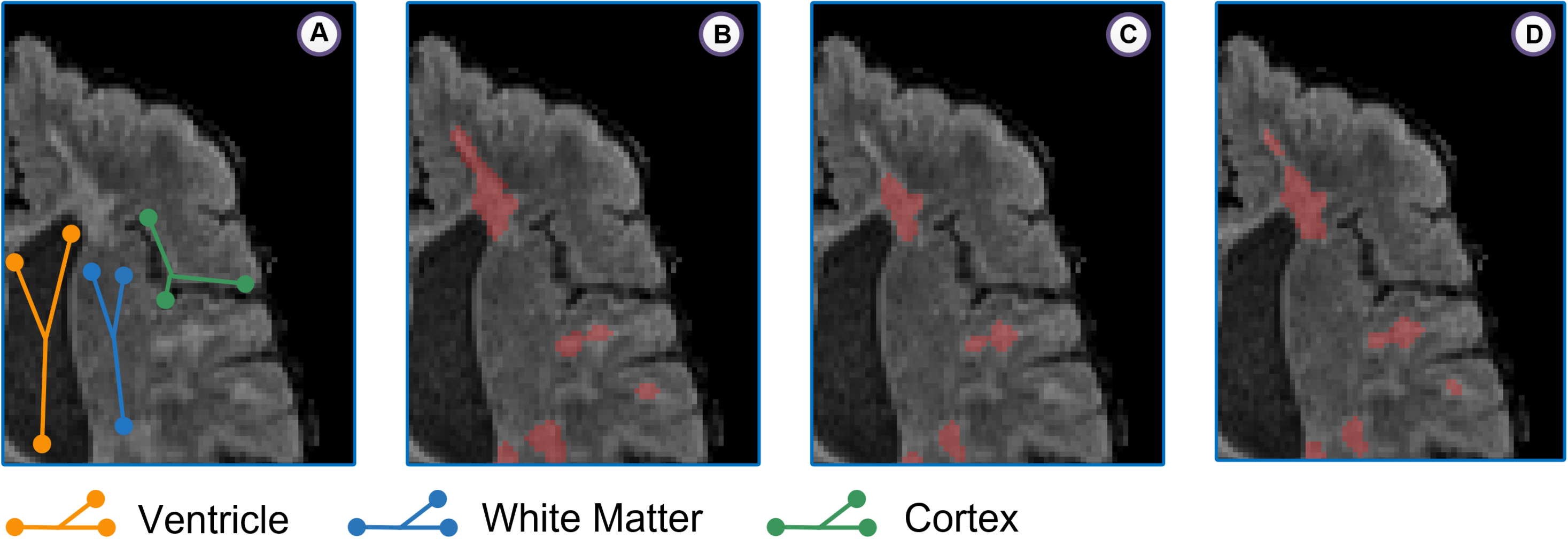}
	\caption{
        Visualization of a case for brain lesion segmentation.
	Colored arrows indicate regions of different tissues.   
	(A) A T2-FLAIR image example of a patient with a heavy lesion burden.
	(B) Lesions labeled by a human expert (marked in red). 
	(C) Segmentation result from baseline network \protect\cite{ZHANG2021102854} without SCP. 
	(D) Segmentation result from baseline network with SCP.
        }
        \label{fig:failure-case}
\end{figure}


In contrast to natural images, a large portion of medical images depict human organs through non-invasive imaging techniques such as magnetic resonance imaging (MRI), and these images exhibit more structured patterns than natural images due to their shared similarities across the cohort.
Taking the human brain as an example, while brains from different subjects differ in size, disease type or other personal characteristics, all these brains have shown similar hierachical tissue structures across the cohort, e,g., the location and shape of ventricles, white matter and cortex (see Fig.~\ref{fig:failure-case}).
However, most of existing studies on lesion segmentation apply CNN architectures that are originally designed for natural image segmentation, leading to the the structural information loss.
From three reproducible observations on brain lesion segmentation we present in the following, we hypothesize that imposing spatial invariance throughout all layers of these CNN architectures is overly restrictive on lesion segmentation tasks with clear tissue structures, and such invariance can be relaxed to improve the computational efficiency and meantime maintain or increase segmentation accuracy to suit needs for resource-limited clinical applications.

\begin{figure}[!t]
	\centering
        \includegraphics[width=\columnwidth]{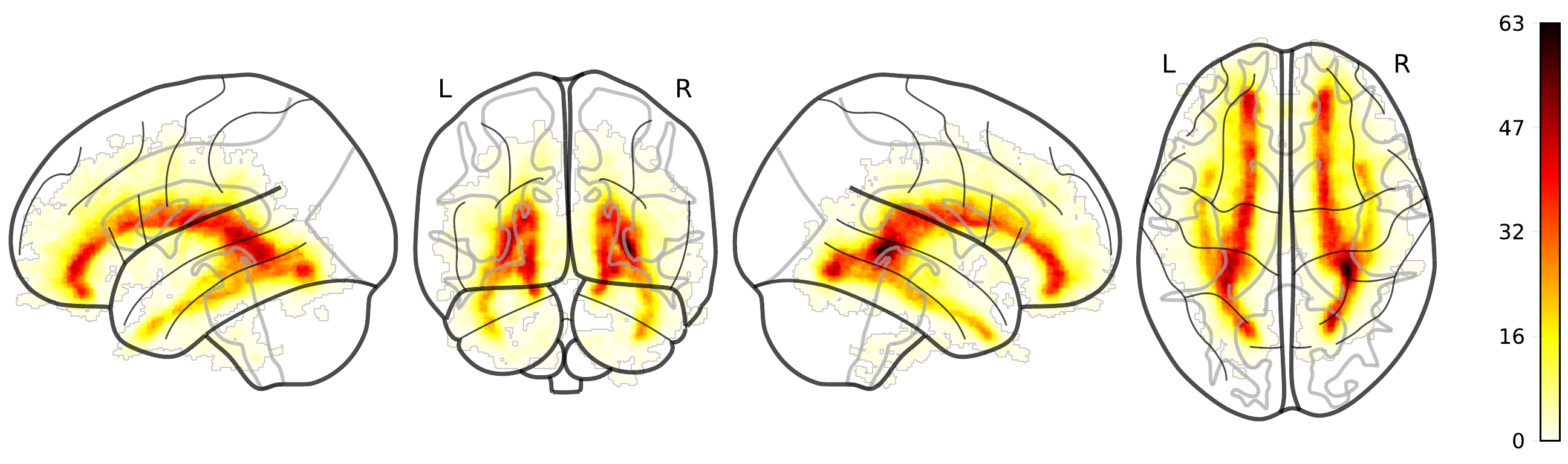}
	\caption{
            A visual example of lesion distribution in human brain. 
            Images and binary lesion masks from a private dataset with 176 subjects are linearly registered to a standard MNI space using FSL neuroimaging toolbox \protect\cite{fischl2012freesurfer}. 
            All registered lesion masks were then summed up to form a unified heat map. The heat map was projected into orthogonal planes as above using \emph{Nilearn} library \protect\cite{abraham2014machine}.
	}
	\label{fig:distribution}
\end{figure}

Our first observation is that, for CNN-based methods, lesions near cortices or ventricles in human brain are more prone to be mis-segmented than those surrounded by the normal-appearing white matter (NAWM) \cite{moll2011multiple} in between ventricles and cortex. 
As shown in Fig.~\ref{fig:failure-case}, lesions inside the white matter (WM) show more consistent image contrast and tissue structures than those near cortices and ventricles.
The change of the context may in part accounts for the mis-segmentation and this raises a question about whether applying the spatially invariant convolution filters is overly restrictive. 
Second, lesions locate diversely among different individuals but the pattern of the lesion density distribution from a patient cohort are rather structured.
As can be seen from Fig.~\ref{fig:distribution}, there is a higher lesion density at the peri-ventricular area of the brain, and roughly the lesion density decreases with the distance from the ventricle.
Third, the effective receptive field of each pixel differs in its location, where pixels closer to the brain center (inner) have a larger receptive field than those farther away (outer), because the outer ones aggregate more zero values than the inner ones due to the existence of a brain boundary.



In general, better performance can be achieved by enlarging the network size to accommodate changes of the tissue context, lesion density distribution, and receptive field; however, simple enlargement of the network without considering domain-specific information may suffer from a higher computational cost or even overfitting due to the limited amount of medical data.
To test our hypothesize, a viable way is replacing the sharing filters with spatially covariant filters.
However, this is not applicable as it adds up a huge amount of computational resources, deviating from our original intention.
Therefore in this paper, we propose spatially covariant pixel-aligned classifier (SCP) to improve the computational efficiency and meantime maintain or increase accuracy for lesion segmentation.
SCP relaxes the spatial invariance constraint imposed by convolutional operations and optimizes an underlying implicit function that maps image coordinates to network weights, the parameters of which are obtained along with the backbone network training and later used for generating network weights to capture spatially variant contextual information. 
SCP is intriguingly simple and can be plugged into most existing segmentation networks.

We apply SCP to two lesion segmentation tasks, white matter hyperintensity (WMH) segmentation in magnetic resonance imaging (MRI) and liver tumor (LiT) segmentation in contrast-enhanced abdominal computerized tomography (CT), to verify our hypothesis.
The main findings of the paper are in three folds: 
1) Relaxing spatial variance for tasks with datasets having structured representation or spatially inhomogeneous features are beneficial.
2) SCP works better for WMH segmentation than LiT segmentation as brains are more structured than liver across different individuals.
3) The experimental results suggest that with SCP relaxing the spatial invariance to a certain degree, 23.8\%, 64.9\% and 74.7\% reduction in GPU memory usage, FLOPs, and network parameter size can be achieved without compromising any segmentation accuracy.


\section{Related Works}

\subsection{Lesion Segmentation Networks}

Lesion segmentation is one of the most important and difficult tasks in clinical applications, because lesions such as WMH and LiT vary greatly in terms of shape, size, and location.
Numerous automated approaches have been proposed to handle the problem, and usually, methods with prior information injected perform better than those using plain neural networks.
The Tiramisu network \cite{zhang2019multiple} uses 2.5D stacked DenseNet \cite{huang2017densely} to capture broader brain structural information, \cite{aslani2019multi} uses multi-branch residual network \cite{he2016deep} to fuse multi-contrast information, and \cite{zhang2019rsanet,zhang2020efficient} applies regularized self-attention networks for aggregation of richer contextual information.
nnU-Net \cite{isensee2021nnu} takes one step further to condense and automate the key decisions for designing a successful segmentation network.
Despite their success, little domain-specific structure for the network has been developed by these methods, which may lead to sub-optimal results.

The distance transformation mapping has been integrated into neural networks through the development of edge-aware loss functions \cite{kervadec2019boundary,zhang2020geometric,karimi2019reducing} and network layers with anatomical coordinates as prior information \cite{ZHANG2021102854}, improving lesion segmentation performance.
These edge-aware loss functions reduces the class imbalance between teh background and lesions, helping segmenting small lesions.
Interestingly, All-Net \cite{ZHANG2021102854} has outperformed its pioneers \cite{zhang2019multiple,aslani2019multi,isensee2021nnu} and achieved the state-of-the-art performance on multiple sclerosis lesion segmentation by explicitly integrating domain-specific information into the neural network.
All these methods have achieved a reasonably good result, and we make one step further to provide the possibility and elasticity to find a better trading-off between the segmentation accuracy and computation efficiency by relaxing spatial invariance.

\subsection{Invariant Representations of CNNs}

While few efforts have been made to study the spatial invariance of CNNs in the context of medical images, many researchers has explored this in natural images.
One line of works seek to learn invariant \cite{yi2016lift,cheng2016learning,liu2019gift} or equivariant representations \cite{cohen2016steerable,esteves2018learning,esteves2018learning,tai2019equivariant} for CNNs. 
The performance gain from these methods is usually accompanied with linearly scaled computational cost. 
Another line of works exploits the hidden information behind the invariant representations learned from CNNs, and our SCP falls in this line.
The most related work to our SCP is a low-rank locally connected (LRLC) layer  \cite{elsayed2020revisiting} that can parametrically adapt the spatial invariance degree for each convolotion layer by controlling its rank parameter; however, it's non-trivial to adapt LRLC for lesion segmentation and there is no apparent way of finding a trading-off between accuracy and computation efficiency.
Interestingly, \cite{kayhan2020translation} empirically shows that CNNs consisting of spatially invariant filters can naturally encode absolute spatial location information, reducing the translation invariance in CNNs, and a removal of such encoding can impose stronger inductive bias on the translation invariance, benefiting image classification tasks with small datasets.
Further, \cite{islam2019much} analyzes how much spatial information CNNs encode and how CNNs encode, particularly the authors show that the zero-padding near the image border is the key to deliver the spatial location information; however, this implicit spatial information comes at the price of using deeper CNNs or larger convolution filters.

\cite{kayhan2020translation} and \cite{islam2019much} show that some part of the CNN capacity is occupied by implicit spatial location encoding, meaning that it is possible to preserve this part of capacity for feature learning by feeding directly the spatial location information such as coordinates \cite{liu2018intriguing} to the network.
In addition, considering the hierachical structure, boundary shape and spatially inhomogeneous contrast features of human brain, our proposed SCP can further exploit these information and reduce computational complexity.

\section{Methodology}


\subsection{Preliminaries}

Let $\mathbf{X} \in \mathbb{R}^{C\times H \times W}$ be an input feature map with $C$ channels ($H$ and $W$ denote the input size, and $C$ is the number of input channels), $\mathbf{X}_{ij}$ be the feature vector at location $(i,j)$, and $\mathbf{Y} \in \mathbb{R}^{N \times W\times H}$ ($N$ is the number of segmentation categories) be the output of the segmentation head of the network, we can then describe the output logit at position $(i,j)$ as follows:
\begin{equation}
    \label{eq:normal-head}
    \mathbf{Y}_{ij} = f(\mathbf{X}_{ij};\mathbf{w}),
\end{equation}
where $\mathbf{w}$ is the trainable weight of the filters for the segmentation head and $f$ transforms the feature vector in every location in an identical way using $\mathbf{w}$.
The spatial invariance refers to sharing the same set of filters across all spatial locations.
Usually a practical implementation of Eq.~\eqref{eq:normal-head} for an encoder-decoder network is a $1\times1$ convolution layer, which can be written as follows:
\begin{equation}
    \label{eq:normal-head-re}
    \mathbf{Y}_{ij} = \mathbf{w}^{\top}\mathbf{X}_{ij},
\end{equation}
where $\mathbf{Y}_{ij} \in \mathbb{R}^{N\times 1\times 1}$, $\mathbf{w} \in \mathbb{R}^{C\times N}$ and $\mathbf{X}_{ij} \in \mathbb{R}^{C\times 1\times 1}$.

\subsection{Spatially Covariant Segmentation Head}
Let $\mathbf{S} \in \mathbb{R}^{M\times H \times W}$ ($M$ is the number of elements for the encoded position vector) be the position encoding tensor, to relax the spatial invariance, we can replace the shared $\mathbf{w}$ with a function of location: 
\begin{equation}
    \label{eq:scp-w}
    \mathbf{W}_{ij}=\Phi(\mathbf{S}_{ij};\theta),
\end{equation}
where $\mathbf{W}_{ij}$ is the generated spatially covariant weight, $\theta$ is the trainable parameter for function $\Phi$.
Basically, function $\Phi$ takes in an encoded position vector $\mathbf{S}_{ij}$ and uses its parameter determined by $\theta$ to generate weights for convolution filters.
With Eq.~\eqref{eq:scp-w}, we can derive the spatially covariant segmentation head as follows:
\begin{equation}
    \label{eq:scp-head}
    \mathbf{Y}_{ij} = g(\mathbf{X}_{ij};\Phi(\mathbf{S}_{ij};\theta)).
\end{equation}
Eq.~\eqref{eq:scp-head} can be seen as a generalization of Eq.~\eqref{eq:normal-head}, as the former reduces to the latter when $g$ equals $f$ and $\Phi(\mathbf{S}_{ij};\theta)$ outputs the $\theta \in \mathbb{R}^{C\times N}$ regardless of the location information.
Thus, by optimizing the function  $\Phi$ to generate spatially covariant weights across all locations, the spatial invariance is relaxed.

\begin{figure}[!t]
	\centering
	\includegraphics[width=1.0\columnwidth,height=0.521\columnwidth]{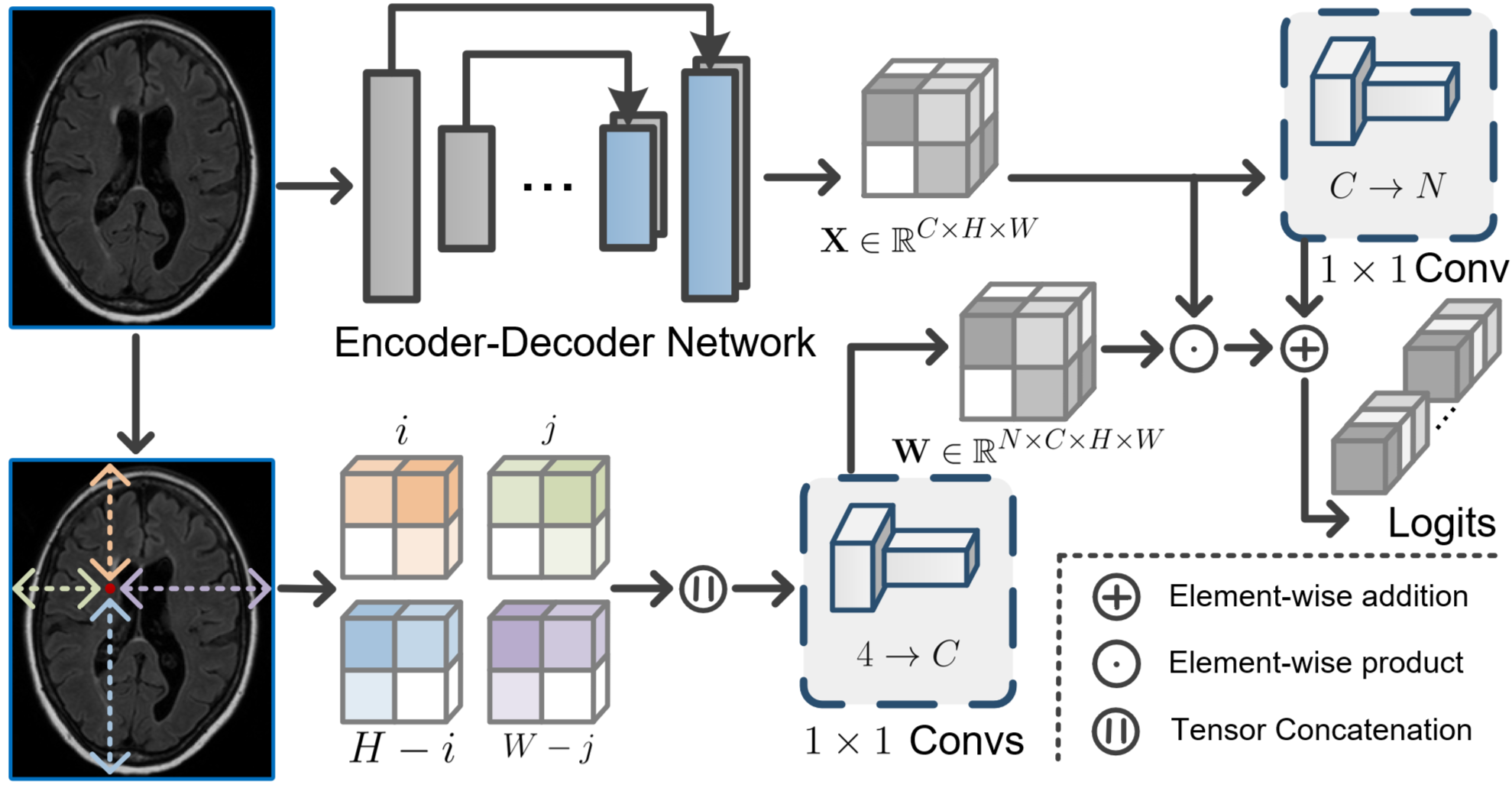}
	\caption{
	Schematic of the proposed SCP.
	}
	\label{fig:framework}
\end{figure}

\subsection{Position Encoding}

Let $\mathbf{X} \in \mathbb{R}^{C\times H \times W}$ be an output feature map with $C$ channels from a encoder-decoder network, we encode the position at $(i,j)$ of $\mathbf{X} $ as follows:
\begin{equation}
    \mathbf{S}_{ij}=(i,j,H-i,W-j).    
\end{equation}
For the rest of the paper, we will refer to $\mathbf{S}_{ij}$ instead of $(i,j)$ as the pixel location.
The Fourier representation \cite{rahaman2019spectral} of the Cartesian coordinates as adopted by neural scene representation \cite{mildenhall2020nerf} or Transformer architecture \cite{vaswani2017attention} is not adopted, because in our application, the surface of generated filter weights is smooth among adjacent pixels, containing no high-frequency information, and enforcing high-frequency variation introduces noise and degrades performance. 
Also, please note that we use four distances (\emph{i.e.} the distance to the left, right, top, and bottom) to represent a pixel coordinate, because the input size $(H,W)$ of the image may vary among different individuals.
Compared to encoding position vector as $\mathbf{S}_{ij}=(i,j,H,W)$, our four-distance encoding method helps expand a smoother surface for the implicit function $\Phi$.

\subsection{Modeling SCP with Neural Networks}

It has been shown that the effective receptive field of a CNN has a Gaussian distribution \cite{luo2016understanding}.
Without loss of generality, we hypothesize that feature vector at each position $\mathbf{S}_{ij}$ follows a multi-variate Gaussian distribution (the dimension is the number of channels $C$) as follows:
\begin{equation}
    \mathbf{X}_{ij} \sim  \mathcal{N}(\mathbf{\mu}_{ij},\mathbf{\Sigma}_{ij}~|~ \mathbf{S}_{ij}),
\end{equation}
where $\mathbf{\mu}_{ij}$ and $\mathbf{\Sigma}_{ij}$ are the corresponding mean vector and covariance matrix in location $\mathbf{S}_{ij}$.
While the normal CNN assumes that $\mathbf{\mu}_{ij}$ and $\mathbf{\Sigma}_{ij}$ stay the same across all spatial locations, we relax this restriction by optimizing an implicit function that maps the pixel location $\mathbf{S}_{ij}$ to the parameter space of $\mathbf{\mu}$ and $\mathbf{\Sigma}$ as follows:
\begin{equation}
    \mathbf{\mu}_{ij},\mathbf{\Sigma}_{ij} = \Phi(\mathbf{S}_{ij};\theta),
\end{equation}
where $\Phi$ represents the implicit mapping function which outputs a continuous parameter space of $\mathbf{\mu}$ and $\mathbf{\Sigma}$ to generate spatial covariant filter weights.
In practice, we use $1\times 1$ convolution layers to approximate this continuous parameter space and optimize the weight $\theta$ of $\Phi$ along with other network weights to map each location $\mathbf{S}_{ij}$ to the corresponding mean vector and covariance matrix.

\begin{figure}[!t]
	\centering
	\includegraphics[width=1.0 \columnwidth,height=0.287\columnwidth]{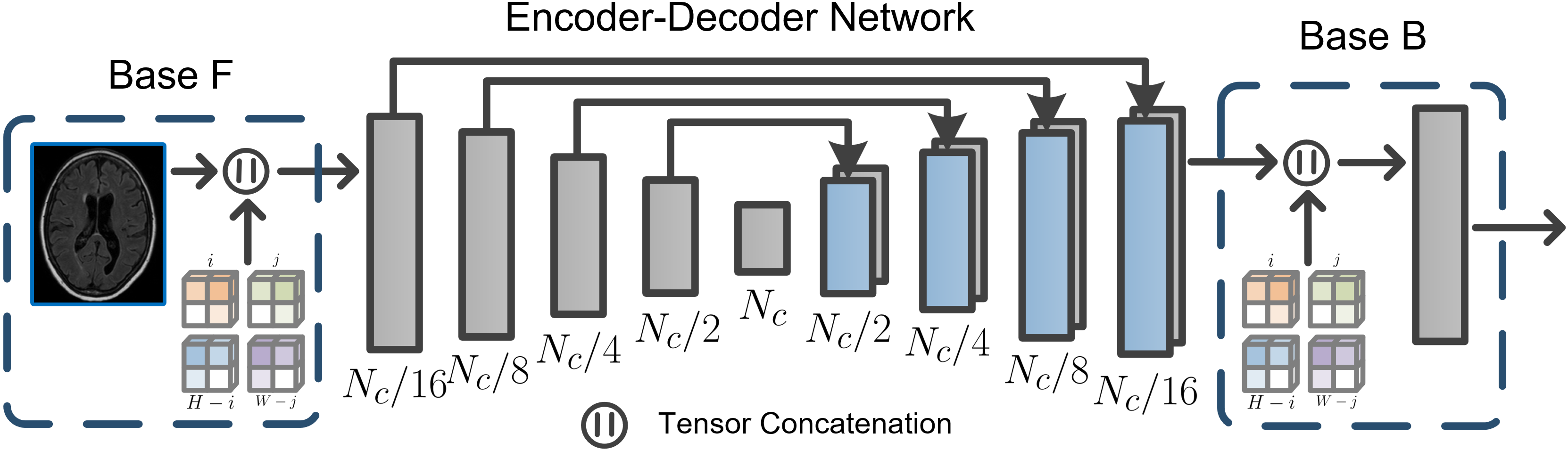}
	\caption{
            Convolutional encoder-decoder backbone network implementing the feature extraction. 
            Each rectangle represents a feature tensor, generated from its preceding feature tensor using a 2D convolution layer. 
            Skip connections are represented by arrows, concatenating encoder and decoder feature tensors.
            The dotted box in the left shows how coordinates are used in Base F models, and the dotted box in the right shows how coordinates are used in Base B models.
	}
	\label{fig:encoder-decoder}
\end{figure}

\subsection{Derivation of SCP}

It has been shown that deep CNNs can linearize \cite{bengio2013better,upchurch2017deep} the manifold of image features into a globally Euclidean subspace of deep features, meaning that the elements of the deep feature vector have minimal interactions with each other. 
As the input feature map for SCP has passed through an entire encoder-decoder network, the output of the implicit function can be further simplified.
More specifically, the covariance matrix $\mathbf{\Sigma}_{ij}$ can be reduced to a variance vector $\mathbf{\sigma}_{ij}$.
We can can then normalize and output the logit of the feature vector at position $\mathbf{S}_{ij}$ as follows:
\begin{align}
\mathbf{\mu}_{ij},\mathbf{\sigma}_{ij} & = \Phi(\mathbf{S}_{ij};\theta), \\
\mathbf{Y}_{ij} & = \mathbf{w}^{\top}\dfrac{\mathbf{X}_{ij} - \mathbf{\mu}_{ij}}{\mathbf{\sigma}_{ij}},
\end{align}
Where $\mathbf{\mu}_{ij} \in \mathbb{R}^{C\times 1\times 1}$ and $\mathbf{\sigma}_{ij} \in \mathbb{R}^{C\times 1\times 1}$ are estimated from the implicit function $\Phi$ with $\theta$ as the trainable parameter, 
$\mathbf{Y}_{ij}$ is the output logit at position $\mathbf{S}_{ij}$, and $\mathbf{w}$ is the weight of the final $1\times 1$ convolution classifier. 
Note that as we use Z-score to normalize the input image, there is no bias term for the classifier.
To be numerically stable, in practice, we estimate $\mathbf{P}_{ij}=1/\mathbf{\sigma}_{ij}$ and $\mathbf{Q}_{ij}=-\mu_{ij}/\mathbf{\sigma}_{ij}$
instead, and the equation for normalization and classification of the feature vectorcan be altered as:
\begin{equation}
\mathbf{Y}_{ij} = \mathbf{w}^{\top} (\mathbf{X}_{ij}\circ\mathbf{P}_{ij} + \mathbf{Q}_{ij}),
\label{eq:musig}
\end{equation}
where $\circ$ is the Hadamard product.
We have derived the spatially covariant segmentation head using the implicit function $\Phi$ in Eq.~\eqref{eq:musig}. 
However, in practice, using Eq.~\eqref{eq:musig} directly may generate unnecessary GPU memory usage due to redundant intermediate computation. 
Looking closer into Eq.~\eqref{eq:musig}, we can reformulate it as:
\begin{equation}
\mathbf{Y}_{ij} = \mathbf{X}_{ij}^{\top}(\mathbf{w}\circ\mathbf{P}_{ij}) + \mathbf{w}^{\top}\mathbf{Q}_{ij},
\label{eq:musig_1}
\end{equation}
where the first term $\mathbf{w}\circ\mathbf{p}_{ij}$ in Eq.~\eqref{eq:musig_1}  can be replaced with a single $\mathbf{W}_{ij}$ generated by the implicit function $\Phi$ directly.
The second term in Eq.~\eqref{eq:musig_2} becomes a scalar, serving as a pixel-wise bias term, but as aforementioned, the bias term is not necessary.
Upon removing the bias term, we add one residual term to stabilize the network training, and our SCP is derived as follows:
\begin{align}
\mathbf{W}_{ij} & = \Phi(\mathbf{S}_{ij};\theta), \\
\mathbf{Y}_{ij} & = (\mathbf{W}_{ij}+\mathbf{w}_r)^{\top}\mathbf{X}_{ij},
\label{eq:musig_2}
\end{align}
where $\mathbf{W}_{ij} \in \mathbb{R}^{N \times C}$ is estimated from the implicit function $\Phi$ with $\theta$ as the trainable parameter, $\mathbf{w}_r\in \mathbb{R}^{N \times C}$ denotes the weight of the residual $1\times 1$ convolution, serving as the shared term for pixels across all spatial locations.

\subsection{Overall Framework Implementation with SCP}

The overall framework consists of an encoder-decoder structured feature extraction network and the proposed SCP module for pixel-wise classification.
As shown in Fig.~\ref{fig:framework}, the input image is passed through an encoder-decoder CNN to obtain the intermediate feature map $X\in\mathbb{R}^{C\times H\times W}$, and in the meantime, pixel-wise position encoding is obtained by concatenating matrices with distances to the left, right, top and bottom of the images.
The spatially covariant weight tensor $\mathbf{W}\in\mathbb{R}^{N\times C\times H\times W}$ is then generated by the implicit function $\phi$ consisting of two layers of a $1\times 1$ convolution followed by a ReLU activation function and a single $1\times 1$ convolution.
With $\mathbf{W}_{ij}$ for each pixel position, we apply Eq.~\eqref{eq:musig_2} to output the spatial variant logits for the feature vector of each position.
The final segmentation can be obtained by the sum of the output from the residual term and the output from the implicit function, and a Sigmoid function.


\begin{table*}[!ht]
\caption{
    Quantitative comparisons of WMH segmentation results produced by the proposed SCP and other network variants. 
    The best performer for each metric within the same complexity is bold-faced.
    All ratios of FLOPs and GPU memory usage are computed using an input tensor with size $(3\times 160 \times 224)$.
}
\label{tab:wmh_table}
\begin{center}
\resizebox{1.9\columnwidth}{!}{
\begin{tabular}{ lccccccccc }
\hline
\hline
 Model   & $N_c$         & Dice (\%) & L-Dice (\%) & L-F1 (\%) & L-PPV (\%) & L-TPR (\%) & Memory Ratio & FLOPs Ratio & Parameter Size Ratio  \\
\hline
Base  & \multirow{4}{*}{\underline{128}} & 67.41      & 51.53       & 57.53      & 58.48   & 56.62   &1.00	&1.00	&1.00  \\
Base F  &  & 27.94      & 24.41       & 31.91      & 45.25      & 24.64  &1.06	&1.02	&1.00 \\
Base B  &  & 40.95      & 31.44       & 38.24      & \bf{90.95} & 24.21 &1.17	&1.02	&1.00 \\
SCP (ours)    &     & \bf{74.47} & \bf{61.30}  & \bf{66.39} & 67.28      & \bf{65.53} &1.78	&1.42	&1.02 \\
\hline
Base   & \multirow{4}{*}{\underline{256}}   & 75.29       & 64.90      & 69.40      & 71.09      & 67.78 &2.13	&3.96	&4.02 \\
Base F  &  & 59.39       & 39.67      & 47.22      & 44.37  	& 50.45  &2.19	&4.00	&4.02\\
Base B  &  & 67.82       & 56.74      & 64.02      & \bf{83.41} & 51.95 &2.43	&4.00	&4.02 \\
SCP (ours)   &     & \bf{77.63}  & \bf{69.55} & \bf{72.27} & 69.78      & \bf{74.95} &3.69	&5.51	&4.06 \\
\hline
Base   & \multirow{4}{*}{\underline{512}}   & 75.39       & 65.61      & 70.50      & 67.74      & \bf{73.50 } &4.84	&15.71	&16.06 \\
Base F  &  & 62.63       & 45.85      & 54.10      & 54.48      & 53.73  &4.90	&15.80	&16.08 \\
Base B  &  & 74.90       & 63.89      & 69.97      & \bf{81.67} & 61.20 &5.39	&15.82	&16.08  \\
SCP (ours)  &     & \bf{77.28}  & \bf{70.34} & \bf{73.89} & 75.89      & 72.00 &7.94	&21.76	&16.22\\
\hline
\hline
\end{tabular}
}
\end{center}
\end{table*}

\section{Experiments and Results}

We use PyTorch \cite{paszke2019pytorch} for all network implementations, and train them with an Nvidia RTX 2080Ti GPU.
We carefully carry out a set of experiments to show both strengths and useful characteristics of the proposed SCP. 
We use MRI and CT datasets, which are the most commonly used modalities for medical imaging, to show that SCP generalizes across different imaging techniques.
Meanwhile, brain lesions and liver tumors, two commonly seen diseases, are used to demonstrate SCP's generalization across different human organs.

\subsubsection{Baselines and Comparators} 
All-Net \cite{ZHANG2021102854} is a recently developed network for brain lesion segmentation, outperforming other highly competitive networks such as nnU-Net \cite{isensee2021nnu}, Tiramisu network \cite{zhang2019multiple} and multi-branch residual network \cite{aslani2019multi} on the lesion segmentation task.
Without loss of generality, we adopt the backbone network from All-Net as baseline network.

To illustrate the effectiveness of SCP utilizing spatially variant information rather than taking additional coordinates, two network variants with coordinate convolution \cite{liu2018intriguing} are further compared: 1) baseline network with coordinates as an additional input channel (denote as Base F); 2)  baseline network with coordinate convolution replacing the final convolution layer (denote as Base B).
We also deploy three variants with different model capacity by varing the number of channels $N_c$ in the encoder-decoder network (see Fig. \ref{fig:encoder-decoder} for more details about how $N_c$ changes the network capacity), in which $N_c$ denotes the maximum number of channels in the center layer of the network (\emph{e.g.} the Base B network with $N_c=256$ is denoted as Base B (256)). 

\subsubsection{Evaluation metrics.} 

To quantify the models' performance, we use the Dice similarity coefficient~\cite{dice1945measures}, lesion-wise Dice (L-Dice), lesion-wise true positive rate (L-TPR), lesion-wise positive predictive value (L-PPV), and lesion-wise F1 score (L-F1) as metrics.
L-Dice, LTPR, and LPPV are defined as $\text{L-Dice} = \frac{\text{TPR}}{\text{GL}+\text{PL}}$, $\text{L-TPR} = \frac{\text{TPR}}{\text{GL}}$, and $\text{L-PPV} = \frac{\text{TPR}}{\text{PL}}$, where TPR denotes the number of lesions in ground-truth segmentation that overlap with a lesion in the produced segmentation, and GL, PL are the numbers of lesions in ground-truth segmentation and produced segmentation, respectively.
L-F1 is the harmonic mean of L-TPR and L-PPV: $\text{L-F1}=\frac{2\text{L-TPR}\times\text{L-PPV}}{\text{L-TPR}+\text{L-PPV}}$.
In general, Dice quantifies the voxel-wise overlap between the output and the ground truth, while L-Dice and L-F1 are more sensitive in measuring the lesion-wise detection accuracy.
In addition, we use the two-tailed paired t-test to compare Dice and L-Dice metrics for the proposed SCP with other network variants to demonstrate statistical significance for the improved computational efficiency.




\subsection{White Matter Hyperintensity Segmentation}

Our first experiment is WMH segmentation in human brain. 
WMH is an important type of brain lesion whose precise tracing can provide useful biomarkers for clinical diagnosis and assessment of disease progression.
We use a publicly available \footnote{\url{https://wmh.isi.uu.nl}} WMH segmentation challenge dataset \cite{kuijf2019standardized} for our experiments.
The dataset contains 60 subjects acquired from different scanners of three institutes.
For each subject, a 3D T1-weighted (T1-w) and a 2D multi-slice fluid-attenuated inversion recovery (FLAIR) image are provided.
All images are bias-corrected using SPM12 software \cite{ashburner2000voxel}, followed by co-registering each T1-w image to the FLAIR image. 
The spatial resolution for each image varies from $0.95\times0.95\times3\ mm^3$ to $1.21\times1\times3\ mm^3$, and manual reference segmentation is provided by the consensus of four observers.
In the experiments, the dataset is randomly split into three subsets for model training (42), validation (6), and testing (12).
The final results of each model are the average of three independent runs with different random seeds (random seed uses the system time and is not sensitive).  

\subsubsection{Implementation details.}

We slice the original volumes into a stack of consecutive images. 
Each volume is center-cropped to a fixed size of $160\times224$ to accommodate batch training, followed by Z-score based intensity normalization.
T1-w and FLAIR images of each subject are concatenated through the channel dimension before being used as the input to the network.
To train the network, we use Adam~\cite{kingma2014adam} optimizer, with an initial learning rate of $10^{-3}$ (weight decay of $10^{-6}$), and a batch size of 14.
The learning rate is halved at 50\%, 70\%, and 90\% of the total training epoch (90) for optimal convergence.

\begin{figure*}[!t]
    \label{fig:ttest}
	\centering
	\subfloat[WMH on Dice.]{\includegraphics[width=.53\columnwidth]{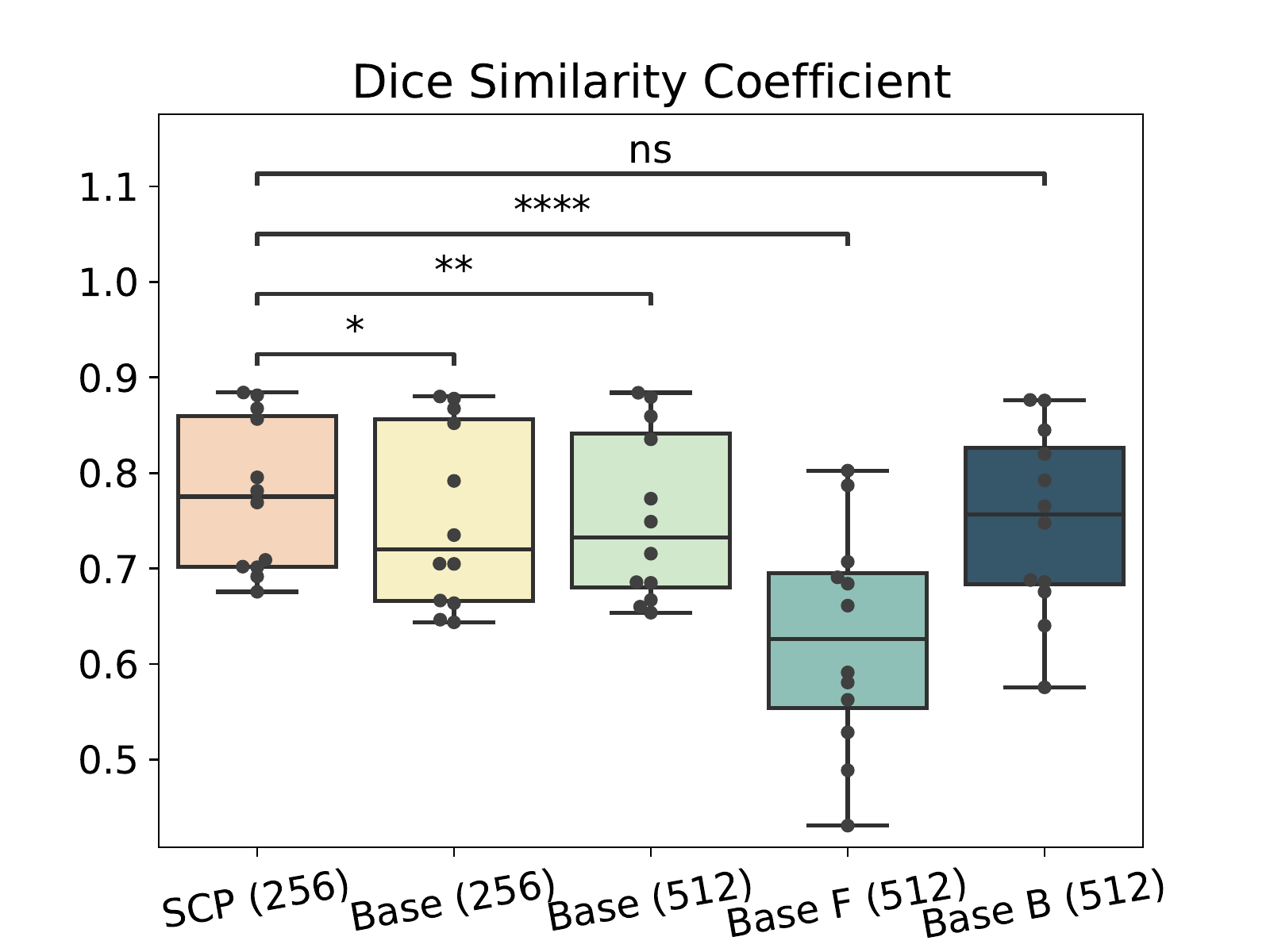} \label{fig:wmh_dice_ttest}}
	\subfloat[WMH on L-Dice.]{\includegraphics[width=.53\columnwidth]{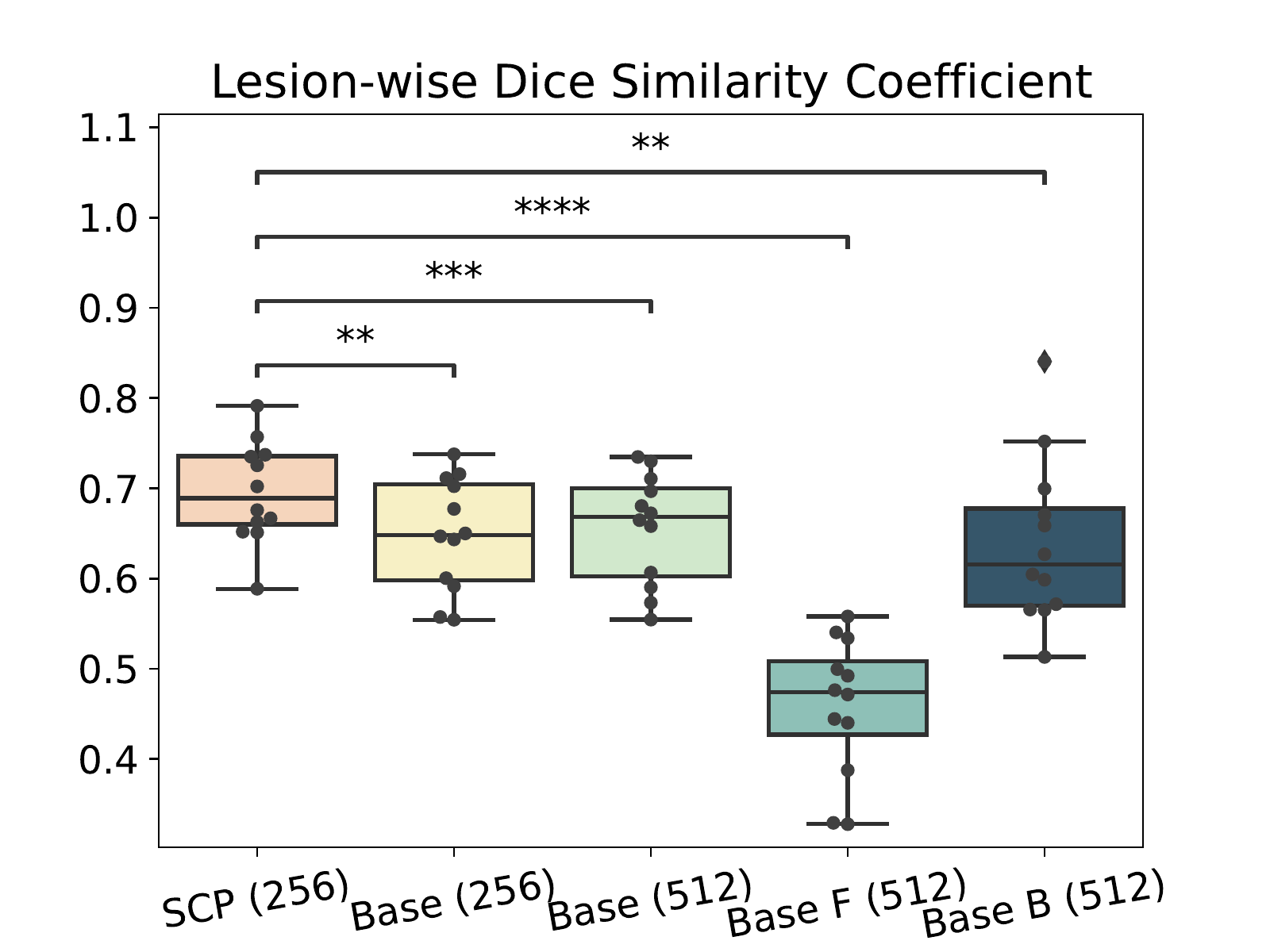} \label{fig:wmh_ldice_ttest}} 
	\subfloat[LiTS on Dice.]{\includegraphics[width=.53\columnwidth]{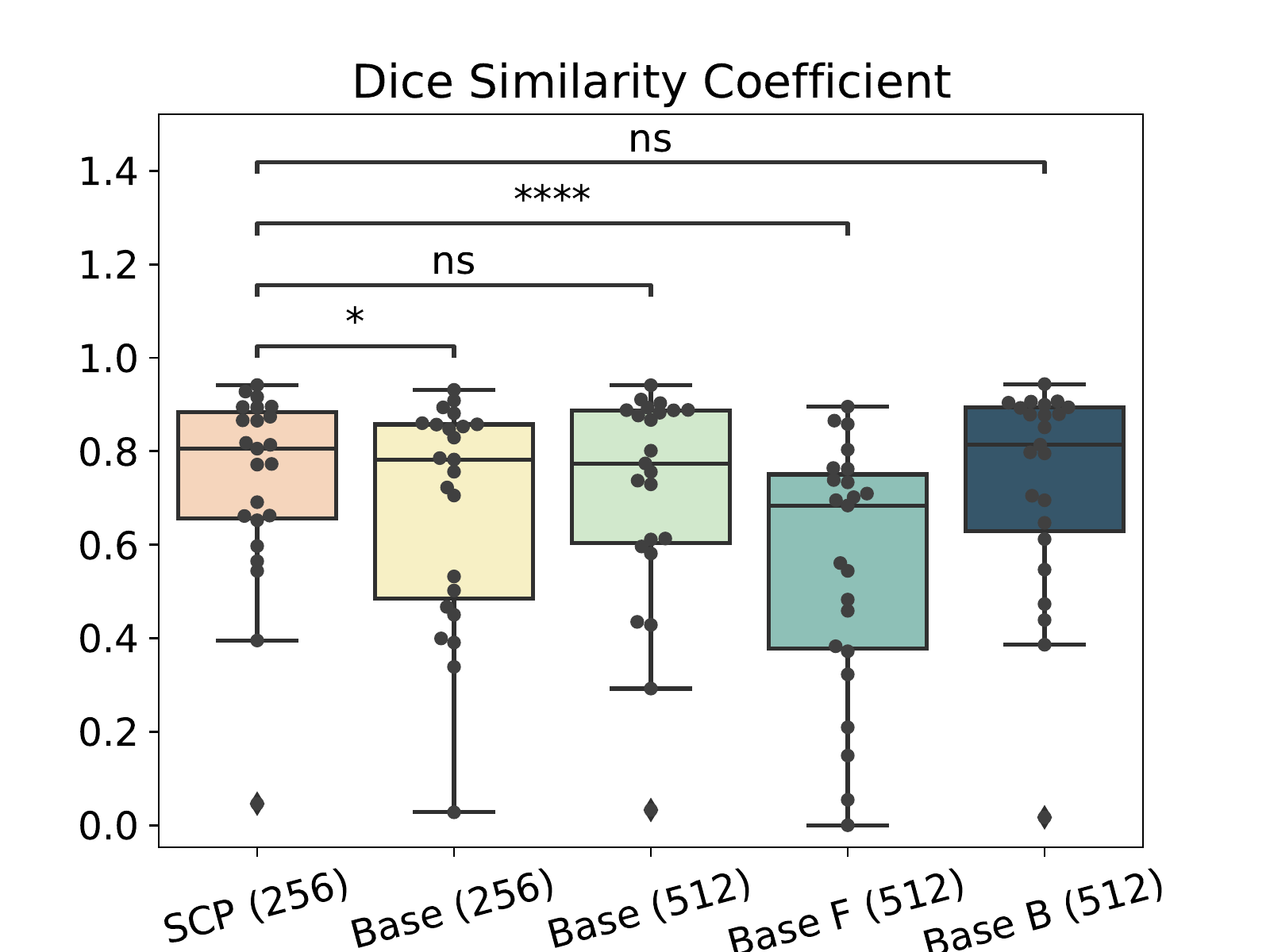} \label{fig:lits_dice_ttest}}
	\subfloat[LiTS on L-Dice.]{\includegraphics[width=.53\columnwidth]{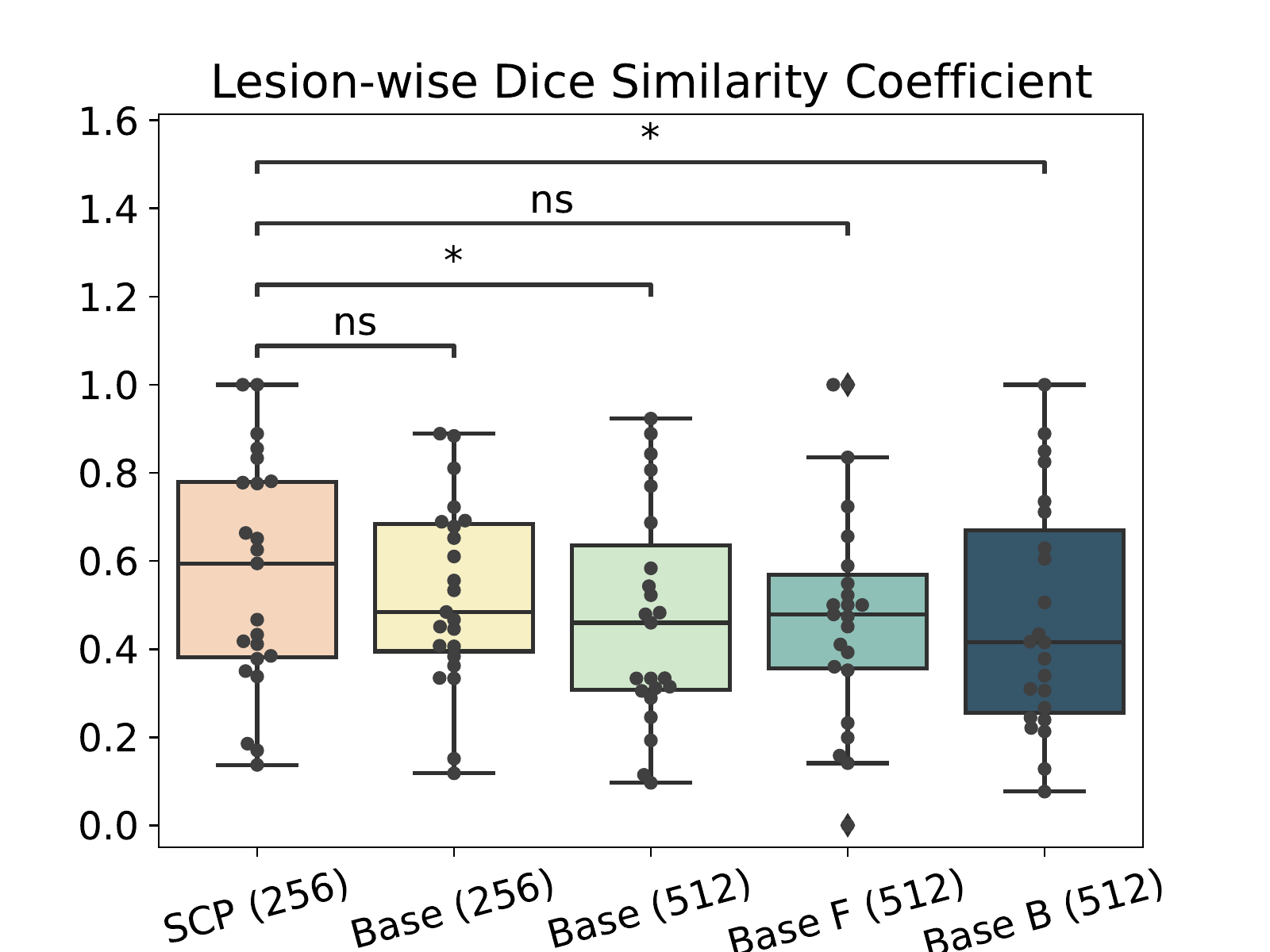} \label{fig:lits_ldice_ttest}}
	\caption{
	Performance comparison of different network architectures with Dice 
	and L-Dice 
	evaluated on the testing set of WMH and LiTS datasets.
	The number in the bracket indicates the number of channels $N_c$ for the network.
	Statistical significance test between of the proposed SCP and other comparators are evaluated using a paired t-test. 
	The threshold of the significance is $\alpha$= 0.05, and the p-values in the figure are annotated as: * for $p < 0.05$, ** for $p < 0.01$, *** for $p < 0.001$, **** for $p < 0.0001$, and ns for non-significant.}
\vspace{-2ex}
\end{figure*}


\subsubsection{Quantitative results.}

We compare the proposed SCP with the baseline network \cite{ZHANG2021102854} and other variants. 
Three different levels of network capacities ($N_c=128$ for low-capacity, $N_c=256$ for moderate-capacity, and $N_c=512$ for high-capacity) of the backbone networks are adopted. 
Pixel-wise and lesion-wise metric scores as well as ratios of floating point operations per second (FLOPs), memory usage and network parameter size between each network variant and the Base (128) network are shown in Table~\ref{tab:wmh_table}. 

In the low-capacity group ($N_c=128$), it is observed that SCP (128) outperforms Base (128), Base F (128), and Base B (128) in all metrics except for L-PPV.
Though Base B (128) has the highest L-PPV (90.95\%), the L-TPR (24.21\%) is much lower than SCP (128), resulting in worse L-Dice. 
Importantly, SCP (128) has at least 7.1\% and 9.8\%  improvement in Dice and L-Dice compared to other network variants with the same compacity.
Similar trends in performance are observed among the moderate- and high-capacity groups ($N_c=256$ or $N_c=512$), where SCP shows a consistent improvement compared to those networks without the proposed method.
Interestingly, both Base F and Base B which take additional coordinate information, perform consistently worse than SCP in all three levels of model capacity, demonstrating the effectiveness of SCP utilizing spatially variant contextual information.
In addition, SCP with moderate-capacity backbones achieves better performance compared to other network variants in a higher-capacity backbone.
This results in 23.8\%, 64.9\% and 74.7\% reduction in GPU memory usage, FLOPs, and network size respectively without sacrificing any segmentation accuracy.


\begin{table}[!h]
\caption{
    Quantitative comparisons of LiTS segmentation results produced by neural networks of three levels of complexities. 
    The best performer for each metric is bold-faced.
}
\label{tab:liver_table}
\vspace{-1ex}
\begin{center}
\resizebox{1.\columnwidth}{!}{
\begin{tabular}{ lcccccc }
\hline
\hline
Model   & $N_c$    & Dice (\%) & L-Dice (\%)    & L-F1 (\%)  &  L-PPV (\%) & L-TPR (\%)    \\
\hline
Base  & \multirow{4}{*}{\underline{128}}   & 55.57      & 45.76        & 67.29      & 61.20      & 74.73  \\
Base F  &  & 55.13      & 48.76        & \bf{68.65}     & \bf{64.84}      & 72.94  \\
Base B  &  & 57.39      & 48.11        & 65.71     & 58.91 & 74.29  \\
SCP (ours)  &    & \bf{61.21} & \bf{50.13}   & 68.03 & 58.90      & \bf{80.50} \\
\hline
Base   & \multirow{4}{*}{\underline{256}}   & 67.59       & 53.60        &  72.58  & 63.46      & 84.77  \\
Base F  &  & 54.08       & 46.78        &  66.44  & 63.24  	& 69.98  \\
Base B  &  & 69.63       & 50.33        & 67.84   & 56.01 & 86.02  \\
SCP (ours)   &     & \bf{72.71}  & \bf{59.70} & \bf{75.11} & \bf{64.86}      & \bf{89.20} \\
\hline
Base   & \multirow{4}{*}{\underline{512}}   & 71.36       & 48.80       & \bf{67.98}     & 54.69      & 89.80 \\
Base F  &  & 56.64       & 48.59        &  65.22  & \bf{55.85}      & 78.39  \\
Base B  &  & 73.33       & \bf{49.15}    &  67.57  & 53.82 & \bf{90.75}  \\
SCP (ours)  &    & \bf{73.64}  & 49.05   & 66.25 & 52.71      & 89.16  \\
\hline
\hline
\end{tabular}
}
\end{center}
\vspace{-1ex}
\end{table}




\subsection{Liver Tumor Segmentation}

The second experiment is liver tumor segmentation using images scanned from contrast-enhanced abdominal computerized tomography.
It is a challenging task as the image contrast is highly heterogeneous and shape of the liver is diffusive.
An open challenge\footnote{\url{https://competitions.codalab.org/competitions/17094}} dataset LiTS~\cite{bilic2019liver} from MICCAI 2017 is adopted in our experiment.
The dataset is collected from six clinical centers around the world using different scanners and acquisition protocols.
There are in total 131 scans, where the in-plane resolution of each scan varies from 0.55 mm to 1.0 mm with a slice spacing ranging from 0.45 mm to 6.0 mm.
The horizontal shape of each 3D scan is 512$\times$512, and the number of slices varies from 74 to 987.
In the image pre-processing step, we truncate the intensity values of each scan to [-200, 250] in order to exclude the non-liver information.
Following previous work \cite{ma2022hyper}, we resize the in-plane image from 512$\times$512 to 256$\times$256.
In the experiments, the dataset is split into three subsets for model training (91), validation (13), and testing (27).
Similar to the WMH segmentation task, we report the average of three independent runs with different random seeds for each model.

\subsubsection{Quantitative results.}

As shown in Table \ref{tab:liver_table}, we compare SCP with other network variants Base, Base F, and Base B using different network capacities ($N_c=128$, $N_c=256$ or $N_c=512$).
Please note that FLOPs, GPU memory usage and network parameter size are excluded in Table \ref{tab:liver_table} because LiTS datasets are trained with the same network as done in WMH. 
In the low-capacity group ($N_c=128$), Base F has a slightly better L-F1 and L-PPV, but the proposed scp improves the Dice and L-Dice scores by 11.0\% and 2.8\% respectively. 
SCP (128) has a 5.7\% increment for Dice, a 4.4\% increment for L-Dice, and a 7.6\% increment for L-TPR compared to the Base (128).
In addition, SCP (256) consistently outperforms other variants from the moderate-capacity group in terms of Dice and L-Dice. 
However, similar performance gain as in the WMH segmentation task is not observed in LiTS segmentation in the high-capacity group.
Further, the scores of L-F1 and L-Dice are lower in the high-capacity group than those in the moderate-capacity group, indicating that these networks may have encountered overfitting problem.

\begin{figure}[!t]
	\centering
	\subfloat{\includegraphics[width=.33\columnwidth]{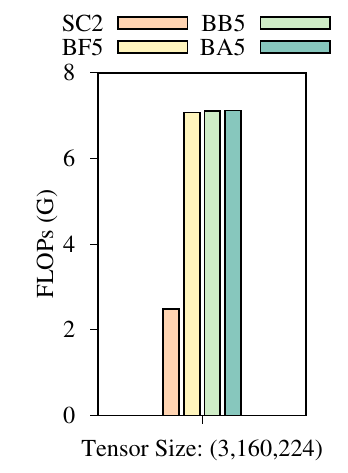} \label{fig:flops}}
	\subfloat{\includegraphics[width=.33\columnwidth]{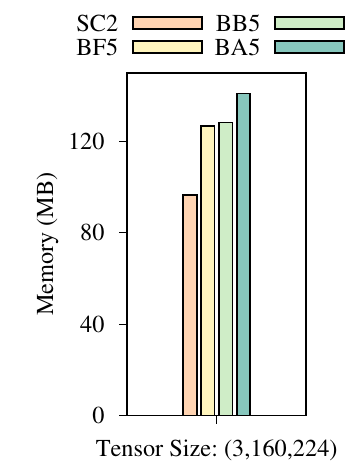} \label{fig:memory}}
	\subfloat{\includegraphics[width=.33\columnwidth]{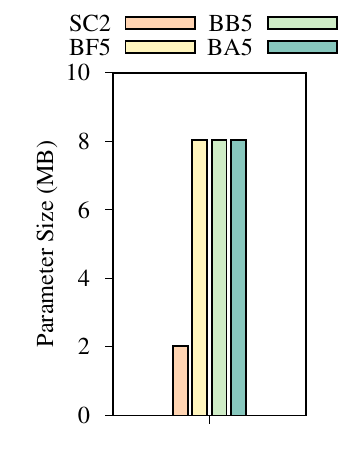} \label{fig:params}}
	\caption{ 
        Computational comparison of GPU memory, FLOPs, and parameter size for four networks.      
        SC2 indicates SCP (256), BF5 indicates Base F (512), BB5 indicates Base B (512), and BA5 indicates Base (512). 
	All the numbers above are obtained from the same machine used for network training. 
        Each module is tested using an input feature tensor with size ($3\times 160 \times 224$).
	}
	\label{fig:complexity}
        \vspace{-1ex}
\end{figure}

\subsection{Statistical \& Complexity Analysis}

Paired t-tests are used compare performance metrics of Dice and L-Dice between the proposed SCP (256) and other network variants Base (256), Base (512), Base F (512), Base B (512).
SCP (256) from the moderate-capacity group is used to compare other variants from the high-capacity group, because the goal for SCP is to improve computational efficiency and in the meantime maintain or achieve better segmentation accuracy by relaxing the spatial invariance.
As we can see from Fig.~\ref{fig:wmh_dice_ttest} and Fig.~\ref{fig:wmh_ldice_ttest}, in WMH segmentation task, except for Base B (512) on Dice, SCP (256) outperforms all other network variants with statistical significance ($p<0.05$) on both Dice and L-Dice.
Fig.~\ref{fig:lits_dice_ttest} and Fig.~\ref{fig:lits_ldice_ttest} show that in LiTS segmentation task, SCP outperforms all other network variants in terms of Dice and L-Dice scores; however, SCP (256) only outperforms Base F (512) on Dice and Base (512) and Base B (512) on L-Dice with statistical significance ($p<0.05$).

Combining results from paired t-tests and Fig.~\ref{fig:complexity}, we can conclude that the proposed SCP (256), relaxing spatial invariance, achieves a significant reduction on computational cost (23.8\%, 64.9\% and 74.7\% reduction in GPU memory usage, FLOPs, and network size) without losing any segmentation accuracy on both WMH and LiTS segmentation tasks.  
The main advantage achieved by SCP is the computation efficiency, favorably making the image analysis tool more accessible in clinical environments without GPU support.

\subsection{Discussions}

\begin{figure}[!t]
	\centering
        \includegraphics[width=\columnwidth]{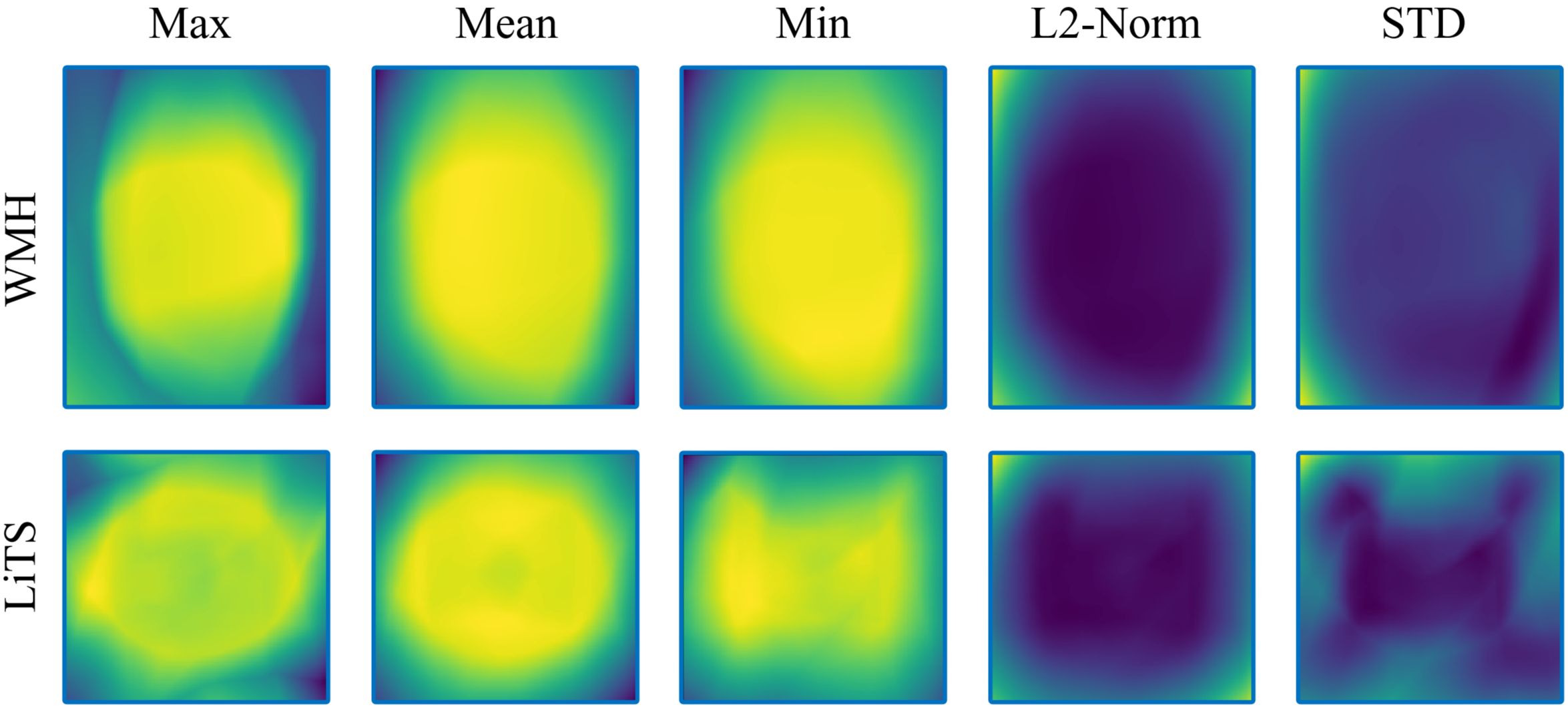}
	\caption{
            A visual example of statistics of the learned spatially covariant weight $\mathbf{W} \in \mathbb{R}^{C \times H \times W}$.
            From left to right are the maximum values, mean values, minimum values, L2-Norm and standard deviation (STD) of $\mathbf{W}$ along the channel dimension. 
	      The first row is from a model trained with WMH datasets, and the second row is from a model trained with LiTS datasets.
            }
	\label{fig:filter_dis}
        \vspace{-1ex}
\end{figure}

To understand how SCP relaxes the spatial invariance and how datasets affect the learning of the spatially covariant weight $\mathbf{W} \in \mathbb{R}^{N\times C \times H \times W}$, we have plotted five key statistics of $\mathbf{W}$ as shown in Fig.~\ref{fig:filter_dis}.
To make value color consistent, we use min-max to normalize each statistic image with value ranging from 0 to 1. 
Please note that as both WMH and LiTS segmentation tasks are binary, $N$ is set to 1.
From left to right in Fig.~\ref{fig:filter_dis} are the maximum values, mean values, minimum values, L2-Norm and standard deviation (STD) of $\mathbf{W} \in \mathbb{R}^{C \times H \times W}$ along the channel dimension respectively. 

As shown in Fig.~\ref{fig:filter_dis}, the weight $\mathbf{W}$ learned from the WMH segmentation shows a stronger pattern than that learned from the LiTS segmentation, where all five statistics of the weight delineate an average image of the shape of human brain.
The statistic images from the LiTS segmentation instead delineate the shape of human breast rather than the target liver.
The boundary of the max value map is clearer in WMH than in LiTS, and the value distribution across five statistics is more consistent in WMH than LiTS.
Moreover, in the max, min and mean value maps from the WMH, value decreases with distance from the center, and especially in the max value map, there is a clear boundary between averaged brain area and background. 
The max value map from WMH segmentation shows consistency with the receptive field where pixels closer to the brain center has a larger receptive field than those farther away.
However, as the shape and location of human liver varies greatly from patient to patient, such phenomenon is not observed from the LiTS segmentation.

Interestingly, we have found out that SCP is more sensitive to small lesions, and more details and qualitative results can be found in the supplementary materials. 
In conclusion, SCP can help relax the spatial variance and benefit lesion segmentation tasks with datasets having structured representation or spatially inhomogeneous features, e.g. aligned human brains with hierachical tissue structure.


\section{Conclusions}

In this paper, we present a novel network module, called spatially covariant pixel-aligned classifier (SCP), to trade-off between computational efficiency and accuracy for lesion segmentation.
SCP relaxes the spatial invariance constraint imposed by convolutional operations and optimizes an underlying implicit function that maps image coordinates to network weights, the parameters of which can be obtained along with the backbone network training and later used for generating network weights to capture spatially variant contextual information. 
We find out that SCP suits lesion segmentation tasks with datasets having structured representation or spatially inhomogeneous features.
We believe that the proposed SCP technique will contribute to more medical imaging applications in the future.

\bibliographystyle{named}
\bibliography{ijcai23}

\newpage

\end{document}